\documentclass[aps,preprint,floats,showpacs,amstex,tightenlines,notitlepage]{revtex4-1}

\usepackage{amsmath}
\usepackage{amsfonts}
\usepackage{amscd}
\usepackage{epsfig}
\usepackage{amssymb}
\usepackage{tabularx}
\usepackage{calligra}
\def\a{\alpha}
\def\b{\beta}

\def\d{\delta}

\def\w{\omega}

\def\p{\partial}

\def\ri{r_{i}}
\def\ro{r_{o}}
\def\tm{{\small {\calligra T }} }

\begin{document}
\title{Unstable fields in Kerr spacetimes}
\author{Gustavo Dotti, Reinaldo J. Gleiser}
\affiliation{Facultad de Matem\'atica, Astronom\'{\i}a y
F\'{\i}sica (FaMAF), Universidad Nacional de C\'ordoba and 
Instituto de F\'{\i}sica Enrique Gaviola, CONICET.\\
 Ciudad
Universitaria, (5000) C\'ordoba,\ Argentina}
%\affiliation{Instituto de F\'{\i}sica Enrique Gaviola, CONICET, C\'ordoba, Argentina.}
\author{Ignacio F. Ranea-Sandoval}
\affiliation{Grupo de Gravitaci\'on, Astrof\'{\i}sica y Cosmolog\'{\i}a, Facultad de Ciencias Astron\'omicas y Geof\'{\i}sicas, Universidad Nacional de La Plata.
 Paseo del Bosque S/N 1900. La Plata, Argentina}

\begin{abstract}
We show that both the interior region $r<M-\sqrt{M^2-a^2}$ of a Kerr
black hole  and   the $a^2>M^2$ Kerr naked singularity admit
unstable solutions of the Teukolsky equation for any value of the
spin weight. For every harmonic number there is at least
one axially symmetric mode   that grows exponentially in time and
decays properly in the radial  directions. These  can be used as
Debye potentials to generate solutions for the scalar, Weyl spinor,
Maxwell and linearized gravity field equations on these backgrounds,
satisfying appropriate spatial boundary conditions and  growing
exponentially in time, as shown in detail for the Maxwell case.
It is suggested that the existence of the unstable  modes is   related to
the so called ``time machine'' region, where the axial Killing
vector field is time-like, and the  Teukolsky equation, restricted
to axially symmetric fields, changes its character from hyperbolic to elliptic.
\end{abstract}


\maketitle

\tableofcontents

\section{Introduction}
 Kerr's solution \cite{kerr} of the vacuum Einstein's equations in Boyer-Lindquist coordinates is
\begin{multline} \label{kerr}ds^2 =  \frac{(\Delta-a^2 \sin^2 \theta) }{\Sigma} dt^2 +2a\sin^2
\theta \frac{(r^2 + a^2 - \Delta)}{\Sigma}
dt d\phi \\- \left[ \frac{(r^2+a^2)^2 - \Delta a^2 \sin^2 \theta}{\Sigma} \right] \sin^2 \theta d\phi^2
-\frac{\Sigma}{\Delta} dr^2 - \Sigma  d\theta^2 ,
\end{multline}
where $\Sigma = r^2 + a^2 \; \cos ^2 \theta$ and $ \Delta = r^2-2Mr
+ a^2$. We  use the metric signature $+---$ to match the formulas in
the original Newman-Penrose null tetrad formulation \cite{np}  and Teukolsky perturbation treatment \cite{teuko}.
Kerr's metric  has two integration constants: the mass $M$,
and the angular momentum per unit mass $a$. They can be obtained as Komar integrals \cite{komar}
using the time translation  Killing vector field  $K^a$ and the axial Killing vector field $\zeta^a$
(in the above coordinates, these are $\p / \p t$ and $\p / \p \phi$ respectively). We will only consider
the case $M>0$, and we will take $a > 0$ without loss of generality,
since for  $a<0$ we can always change coordinates $\phi \to -\phi$,
under which $a \to -a$. If $0<a < M$ (sub-extreme case), the
$\Sigma=0$  ring curvature singularity at $r=0, \theta=\pi/2$ is
hidden behind the black hole  inner and outer horizons  located at
the zeroes of $\Delta$:  $\ri   =  M - \sqrt{M^2-a^2}$ and $\ro   =
M + \sqrt{M^2-a^2}$. As is well known, $\ri$ and $\ro$ are just
coordinate singularities  in (\ref{kerr}), Kerr's space-time can
be extended  through these horizons and new regions isometric to I:
$r > \ro$, II: $\ri < r < \ro$ and III: $r < \ri$ arise ad infinitum
and give rise to the Penrose diagram displayed in Figure 1. In the
extreme case $M=a$, $\ri=\ro$ and region II is absent, however, we
will still call region I (III) that for wich $r>\ri=\ro$
($r<\ri=\ro$). In the ``super-extreme'' case $a>M$ there is no
horizon at all, the ring singularity being causally connected to
future null infinity. This is not a black hole, but a naked singularity.\\

\begin{figure}[h]
\centerline{\includegraphics[height=9cm]{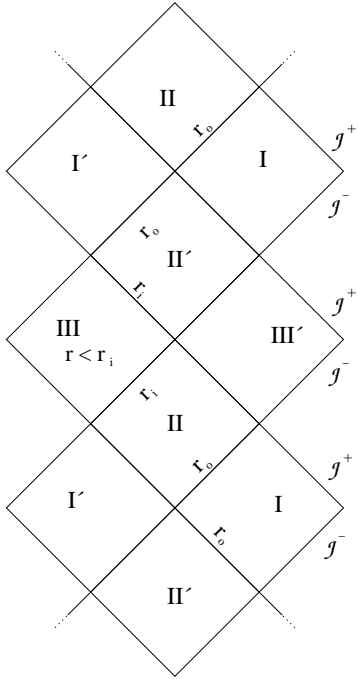}}
\caption{Penrose diagram for the maximal analytic extension of Kerr's space-time. Regions labeled
I and  I' are isometric, and so are II and  II', and III , III'}
\end{figure}

In this paper we study the Kerr naked singularity, and region III of
(sub-extreme and extreme) Kerr black holes. We will refer
to these solutions of the vacuum Einstein equations as KIII and KNS
from now on.
These have a number of
undesirable properties, among which we mention: i) the timelike curvature
singularity as we approach the ring boundary at $r=0,\;
\theta=\pi/2$; ii) the fact that {\em any} two events can be
connected with a future timelike curve
(in particular,
there are closed timelike curves through any point), making KIII and KNS  ``totally
vicious sets'' in the terminology of \cite{onil}, and causing a number
of puzzling causality problems \cite{cp}; iii) the violation of cosmic censorship.\\

Two conjectures are known under the name of cosmic censorship: the
{\em weak} cosmic censorship conjecture establishes that the
collapse of {\em ordinary} matter cannot {\em generically} lead to a
naked singularity, while the {\em strong} cosmic censorship is the
assertion that the maximal development of data given on a Cauchy
surface cannot be {\em generically} continued in a smooth way \cite{qcc}. The
words in italics above signal aspects of the conjectures that need
to be properly specified to turn them into a well defined statement.
In any case, KNS violates  weak cosmic
censorship, and KIII violates strong
cosmic censorship, since it is a smooth extension of the
development of an initial data surface extending from  spatial infinity of region
I to spatial infinity of region I' in  Figure 1,
whose Cauchy horizon agrees with the inner horizon at $\ri$.\\

According to the Carter-Robinson theorem \cite{unique} and further results by
Hawking and Wald, \cite{mazur} if $({\cal M}, g_{ab})$ is an asymptotically-flat
stationary vacuum black hole  that is non-singular on and outside an
event horizon, then it must be  an $a^2  < M^2$  member of the
two-parameter Kerr family.
%In other words, the metric (\ref{kerr}) with $a^2<M^2$ is the  {\em unique} rotating black hole solution of
%Einstein's equations.
The spacetime outside a black hole formed by gravitational  collapse of
 a star is, independently of the characteristics of the collapsing body, modeled by
region I of a Kerr black hole solution, placing this among the  most important of the
known exact solutions of Einstein's equations.
Since this region is stationary (outside the ergosphere), there is a
well defined notion of modal linear stability, under which it has
been shown to be stable \cite{teuko,stable}. It is our opinion,
however, that proving the instability of those stationary solutions of Einstein's equations that display
undesirable features, is as relevant as proving the stability of the physically interesting
stationary solutions. In this line of thought, we are carrying out a
program to analyze the linear stability of the most salient naked
singularities ($M<0$  Schwarzschild spacetime, \cite{dottigleiser} $Q^2 > M^2$
Reissner-Nordstr\"om spacetime \cite{dgrn}, and $a^2>M^2$ Kerr spacetime \cite{doglepu,dgrv}; and also
of those regions lying beyond the Cauchy horizons in
Reissner-Nordstr\"om and Kerr black holes \cite{dgrv,dgrn}. As is well known, the
linear perturbations  of the spherically symmetric Einstein-Maxwell
spacetimes are much easier to deal with than those of the axially
symmetric Kerr spacetime, for which the only separable equations
known to date are not directly related to the metric perturbation.
They are equations satisfied by perturbations of the  components of the Weyl tensor
\begin{equation} \label{weyls}
\psi_0 := -C_{abcd} l^am^b l^c m^d \;\;\;\; \psi_4 := -C_{abcd} n^a \bar m^b n^c \bar m^d
\end{equation}
along a complex null tetrad $l^a, n^a,  m^a, \bar m^a$, among which the only non zero dot products are \cite{np}
\begin{equation} \label{dp}
l^a n_a = 1 , \;\; m^a \bar m _a = -1.
\end{equation}
We use a bar for complex conjugation, $l^a$ and
$n^a$ are real vector fields, whereas $m^a$ is complex. The null tetrad we use
is that introduced by Kinnersley \cite{kin}, given in equation (4.4) in \cite{teuko}.
If we take
\begin{equation}
\varepsilon_{a b c d} =  i \, 4! \; l_{[a}n_{b}m_{c}\bar m _{d]}
\end{equation}
as a right handed volume element, we find that the following two-forms are self dual
\begin{equation}\label{sdb}
\bar m _{[a}n_{b]}, \;\; n_{[a} l_{b]} + m_{[a} \bar m_{b]}, \;\; l_{[a} m_{b]}
\end{equation}
A complex electromagnetic field can be written as
\begin{multline}
F_{ab} := 2 \phi_1 (  n_{[a} l_{b]} + m_{[a} \bar m_{b]} ) + 2 \phi_2  \; l_{[a} m_{b]}
+ 2 \phi_0 \; \bar m _{[a}n_{b]} \\ + 2 \tilde \phi_1 (  n_{[a} l_{b]} + \bar m_{[a}  m_{b]} ) + 2 \tilde \phi_2  \;
 l_{[a} \bar m_{b]}
+ 2 \tilde \phi_0 \;  m _{[a}n_{b]}.
\end{multline}
If $F_{ab}$ is real then $\tilde \phi_j = \bar \phi_j$, if $F_{a b}$ is self dual (anti-self-dual) then
the $\tilde \phi_j$ ($\phi_j$) vanish.
Teukolsky \cite{teuko} found that Maxwell, (Weyl) spinor and scalar fields on a Kerr background can be treated in a similar
way as the gravitational perturbations $\d \psi_0$ and $\d \psi_4$.
 If we take the  components of the  Maxwell fields  (see  (\ref{dp}))
\begin{equation} \label{max}
\phi_0 = F_{ab} l^a m^b \;\;\;\; \phi_1= \frac{1}{2} F_{ab} (l^a n^b + \bar m^a m^b) \;\;\;\;
\phi_2 = F_{ab} \bar m^a n^b ,
\end{equation}
and those of the  two component spinors $\chi_A$
\begin{equation} \label{spin}
\chi_0 = \chi_A o^A \;\;\;\; \chi_1 = \chi_A \iota^A ,
\end{equation}
and weight them with an appropriate power of  the spin coefficient
 $\rho= m^a \bar m^b  \nabla_b l_a = (i a \cos \theta -r)^{-1}$:
\begin{equation} \label{weighted}
\Psi_{\frac{1}{2}} := \chi_0, \;\; \Psi_{-\frac{1}{2}}:=\rho^{-1} \chi_1, \;\; \Psi_1:=\phi_0, \;\;
\Psi_{-1}:= \rho^{-2} \phi_2, \;\; \Psi_2:= \d \psi_0, \;\; \Psi_{-2}:=\rho^{-4} \d \psi_4 ,
\end{equation}
then the (source-free) Maxwell, spinor, and linerized gravity equations can all be encoded
in Teukolsky's master equation \cite{teuko}
\begin{multline} \label{teukopde}
T_s[\Psi_s]:= \left[ \frac{(r^2+a^2)^2}{\Delta}-a^2 \sin^2 \theta \right] \frac{\p^2 \Psi_s}{\p t^2} +
\frac{4Mar}{\Delta} \frac{\p^2 \Psi_s}{\p t \p \phi}
+ \left[ \frac{a^2}{\Delta} - \frac{1}{\sin^2 \theta} \right] \frac{\p^2 \Psi_s}{\p \phi^2} \\ - \Delta^{-s}
\frac{\p}{\p r} \left( \Delta^{s+1} \frac{\p \Psi_s}{\p r} \right) - \frac{1}{\sin \theta} \frac{\p}{\p \theta}
\left( \sin \theta \frac{\p \Psi_s}{\p \theta} \right) -2s \left[ \frac{a(r-M)}{\Delta} + \frac{i \cos \theta}{\sin^2 \theta}
\right] \frac{\p \Psi_s}{\p \phi}\\ -2s \left[ \frac{M(r^2-a^2)}{\Delta}-r-ia \cos \theta \right] \frac{\p \Psi_s}{\p t}
+(s^2 \cot^2 \theta-s) \Psi_s =0.
\end{multline}
The index $s$ in $\Psi_s$ gives the spin weight under tetrad rotations.
The above equation also gives  the massless scalar field equation $\Box \Psi_0=0$
if we set  $s=0$.\\

In \cite{doglepu} we
 found   numerical evidence
that there are solutions of the $s=-2$ Teukolsky equation in the KNS
that grow exponentially in time while satisfying appropriate boundary conditions.
 In \cite{dgrv} we confirmed this fact by proving that
there are {\em infinitely many}  axially symmetric unstable
(meaning, behaving as $e^{kt}$ for some positive $k$)
solutions
of the $s=-2$ equation in the KNS, and also in KIII. \\

In this paper we extend further this result to other linear fields. We show that there are infinitely many
unstable  solutions of the Teukolsky equation
for any $s$ value ($|s|=0,1/2,1,2$). The proof is given in Section \ref{proof}, with some calculations relegated
to the Appendix. The existence of the unstable  modes is shown in Section \ref{stm} to be related to the time machine region
near the ring singularity, which produces a change of character of the Teukolsky PDE -restricted to
axial modes- from hyperbolic to elliptic.
It is suggested that the  emergency of an instability when a PDE changes from hyperbolic to elliptic is generic,
and this is illustrated with a  simple toy model in $1+1$ dimensions. \\

The use of unstable solutions
of Teukolsky's equation as ``Debye'' potentials \cite{reco1}
for constructing unstable spinor, Maxwell and linear gravitational
fields is  illustrated in Section \ref{rec}, where the reconstruction process is explained in  detail
 for Maxwell fields.
All the reconstructed  fields decay properly along spatial directions whilst growing exponentially in time.
This is the notion of instability used in this work. The lack of
a sensible  initial value formulation due to the fact that there are no  partial Cauchy surfaces in KIII and KNS,
 forbids a more traditional approach to the stability issue. This  is quite  different from
 what happens for the Reissner-Nordstr\"om and negative mass Schwarzschild timelike naked singularities, for which
a unique evolution of data given on a partial Cauchy surface can be defined, and instability
proven afterwards \cite{dottigleiser,dgrn}.\\
The following  section contains   information on the Teukolsky equation that is used in the
proof of existence of unstable modes.

\section{Separated Teukolsky equations \label{te}}

Introducing
\begin{equation} \label{sep}
\Psi_s=
R_{\w,m,s}(r) S^m_{\w,s}(\theta) \exp(im\phi) \exp(-i\omega t),
\end{equation}
 the linearized Teukolsky PDE (\ref{teukopde}) is reduced to  a coupled system for $S$ and
$R$,
\begin{equation} \label{ta}
{1\over \sin\theta} {d \over d\theta}\left(\sin\theta {d S\over
d\theta}\right)+\left(a^2\omega^2\cos^2\theta-2 a \omega s \cos\theta -{(m + s \cos\theta)^2\over
\sin^2\theta}  +E -s^2\right)S = 0
\end{equation}
\begin{multline} \label{tr}
\Delta {d^2 R \over dr^2} +(s+1) {d\Delta\over dr} \;{dR\over dr}\\ +\left\{
{K^2-2is(r-M)K\over \Delta}+4ir\omega s -[E-2am\omega+a^2
\omega^2-s(s+1)]\right\}R =0 ,
\end{multline}
where $K=(r^2+a^2)\omega -am$. The system (\ref{ta})-(\ref{tr}) is coupled by
  their common eigenvalue $E$, whose relation
with the separation constant $A$ in \cite{teuko,bertilong} is given by $A=E-s(s+1)$.
Suppose $s$, $m$ and  $\w$ are given, then $E$ in (\ref{ta}) has to be chosen so that $S$ is regular on the sphere.
This gives a denumerable set of eigenvalues that we label as  $E_{\ell}^{ang}(s,m,a \w)$, with $\ell=0,1,2,...$
and $E$ increasing with $\ell$.
In a similar way, $E$ in (\ref{tr}) is chosen such that $R$ decays properly as $|r| \to \infty$
for the KNS ($r \to -\infty$ and $r \to \ri^-$ for KIII), and this also
gives a denumerable set of increasing values $E_n^{rad}(s,m,a\w), n=0,1,2,...$.
A solution of the system (\ref{ta})-(\ref{tr}) is obtained whenever  $E_{\ell}^{ang}(s,m,a \w)
= E_n^{rad}(s,m,a\w) =:E$. Thus, for given $(s,m)$, we may  regard a solution as an intersection
of the curves $E_{\ell}^{ang}$ vs $a \w$ and  $E_n^{rad}$ vs $a \w$, the allowed frequencies
being those at which the curves intersect. This point of view is the one used in the proof of instability below,
for which we restrict our search to $a \w =ik$,  $k$ a real positive number
 (so that (\ref{sep}) gives an $\exp(kt/a)$ behaviour),
 and show that there are intersections for every
$\ell$ and, at least, the fundamental radial mode $n=0$.\\

One way of finding the radial and angular spectra consists in reducing the regularity conditions
 to a continued fraction equation involving $E$ \cite{leaver}. This equation  arises when
 solving a three term recursion relation on the coefficients of a series solution for
  $S$ and $R$ in (\ref{ta}) and (\ref{tr}) \cite{leaver}. An alternative way to obtain the angular spectrum,
which is well suited to the
case we are  interested,  $m=0$ and $a\w=ik$,  is discussed in Section 2.1 of \cite{dgrv}. This approach
is used in the numerical computations leading to Figure 2 below.

\subsection{Angular equation: spin weighted spheroidal harmonics}
 The solutions of  (\ref{ta}) which are regular  on the sphere are called {\em spin weighted spheroidal harmonics}
(SWSH). The spectrum of $E$ values is discrete, and we will use the notation
$E^{ang}_{\ell}(s,m,a \w),$
$\ell=0,1,2,...$ to label the eigenvalues  in increasing order for
a  given set of parameters $(s,m,a\w)$
(the notation is not unified in the literature, note that we use  $\ell =0$
for the lowest eingenvalue  {\em independently} of the spin weight $s$).
Equation (\ref{ta}) exhibits some interesting symmetries:
if  $(S(\theta), E)$ is a regular solution of (\ref{ta}) for some given $(s,m,\w)$ values,
 it is easy to check that $(S(\pi-\theta), E)$ is a solution for $(s,-m,-\w)$ {\em and} for $(-s,m,\w)$,
and that  $(\overline{S(\theta)}, \overline{E})$ is a solution for $(s,m,\bar \w)$. This implies that
\begin{equation} \label{sims}
E^{ang}_{\ell}(s,m,a \w) = E^{ang}_{\ell}(s,-m,-a\w)=E^{ang}_{\ell}(-s,m,a \w) = \overline{E^{ang}_{\ell}(s,m,a \bar \w)}.
\end{equation}

We are interested in  axially symmetric ($m=0$) solutions with
purely imaginary frequencies   ($a\w =ik, k \in {\mathbb R}$). In this case,
from a solution $S(\theta)$ we can extract solutions with real and imaginary
parts of opposite parities by taking  the linear combinations $S(\theta) \pm \overline{S(\pi-\theta)}$.
Also, the eigenvalues  are real, as follows from (\ref{sims}):
\begin{equation}
E^{ang}_{\ell}(s,m=0,ik) = \overline{E^{ang}_{\ell}(s,m=0,ik)}.
\end{equation}
The behaviour of $E^{ang}_{\ell}(s,m=0,a\w=ik)$ in the limits $k \to 0^+$ and $k \to \infty$  will be
relevant in what follows, since, as explained above, an intersection of the curve $E^{ang}_{\ell}(s,m=0,a\w=ik),
k \in {\mathbb R}^+$ with any of the radial curves implies the existence of an unstable (i.e.,
$\sim \exp (kt/a)$) mode.
To study these limits we  will assume that $s \geq 0$ without loss of generality (see (\ref{sims})).
Setting $x:=\cos(\theta), m=0,a\w=ik$ in (\ref{ta}) we find that, for any $k$,
 this equation has regular singular points at $x=\pm 1$,
the possible behaviour of local solutions  around these points being $(1-|x|)^{s/2}$ or $(1-|x|)^{-s/2}$. \\

For $k=0$,
we  expand the regular solutions  as
\begin{equation}\label{eq5}
S(x) = (1-x^2)^{(s/2)} \sum_{j=0}^{\infty} a_j (1-x)^j ,
\end{equation}
and  find that (\ref{ta}) implies
the recursion relation
\begin{equation}\label{eq6}
2 (j+1)(j+1+s)a_{j+1}-\left((j+s)(j+s+1)-E \right)a_j =0,
\end{equation}
which, for large $j$,  gives
  $a_{j+1} \sim a_j /2$. The series in (\ref{eq5}) will thus diverge at $x = -1$ unless we cut
it to down to a polynomial of degree $\ell=0,1,..$ by choosing $E= (\ell+s)(\ell+s+1)$ for
some $\ell=0,1,2,...$. Repeating the calculation for negative $s$, or just using (\ref{sims}), we obtain
\begin{equation}\label{abajoberti1}
E = E_{\ell}^{ang}(s,m=0,a \w=0)= (\ell+|s|)(\ell+|s|+1).
\end{equation}
A more detailed analysis of (\ref{ta}) using continued fraction techniques  gives
 a Taylor expansion for $E^{ang}_{\ell}(s,m,a \w)$ for complex $\w$ near $\w=0$
The expansion up to order $(a \w)^6$ is available in the literature (see
 (\cite{bertilong,seidel} and references therein),
 exhibits the symmetries (\ref{sims}) and has
(\ref{abajoberti1}) as the leading order term, i.e.,
\begin{equation}\label{abajoberti}
E = E_{\ell}^{ang}(s,m=0,a\w)= (\ell+|s|)(\ell+|s|+1) + {\cal O} (\w).
\end{equation}

 Asymptotic expansions
 for $a\w =ik, k \to \infty$ can be found in \cite{lnp,brw,bertishort,bertilong}.
 Particularly useful to our purposes is  that, in our notation
  \cite{bertilong},
 \begin{equation} \label{arrivaberti}
 E^{ang}_{\ell}(s, m, a\w=ik) = (2 \ell  + 1) k + {\cal O} (k^0), \;\;\; \text{ as }  k \to \infty
 \end{equation}
We should warn the reader, however, that a complete proof of the above formula
is not available for $s \neq 0$. Although arguments suggesting the validity of (\ref{arrivaberti}) are given in
\cite{bertilong}, where the formula was also numerically checked for $s=1,2$
the case  $s=1/2$ has not been reported as tested there.
Given that  these equations
are key in the proof of instability for spinor fields, we tested numerically their validity
 using the method developed
in Section 2.1 of \cite{dgrv}.
 We have found and excellent agreement with
both (\ref{abajoberti1}) and (\ref{arrivaberti}) for $s=0,1/2,1,2$.
As an illustration, we give in Figure 2 the results for the  only case not dealt with in the literature, that of $s=1/2$.
\begin{figure}
\centerline{\includegraphics[height=7cm]{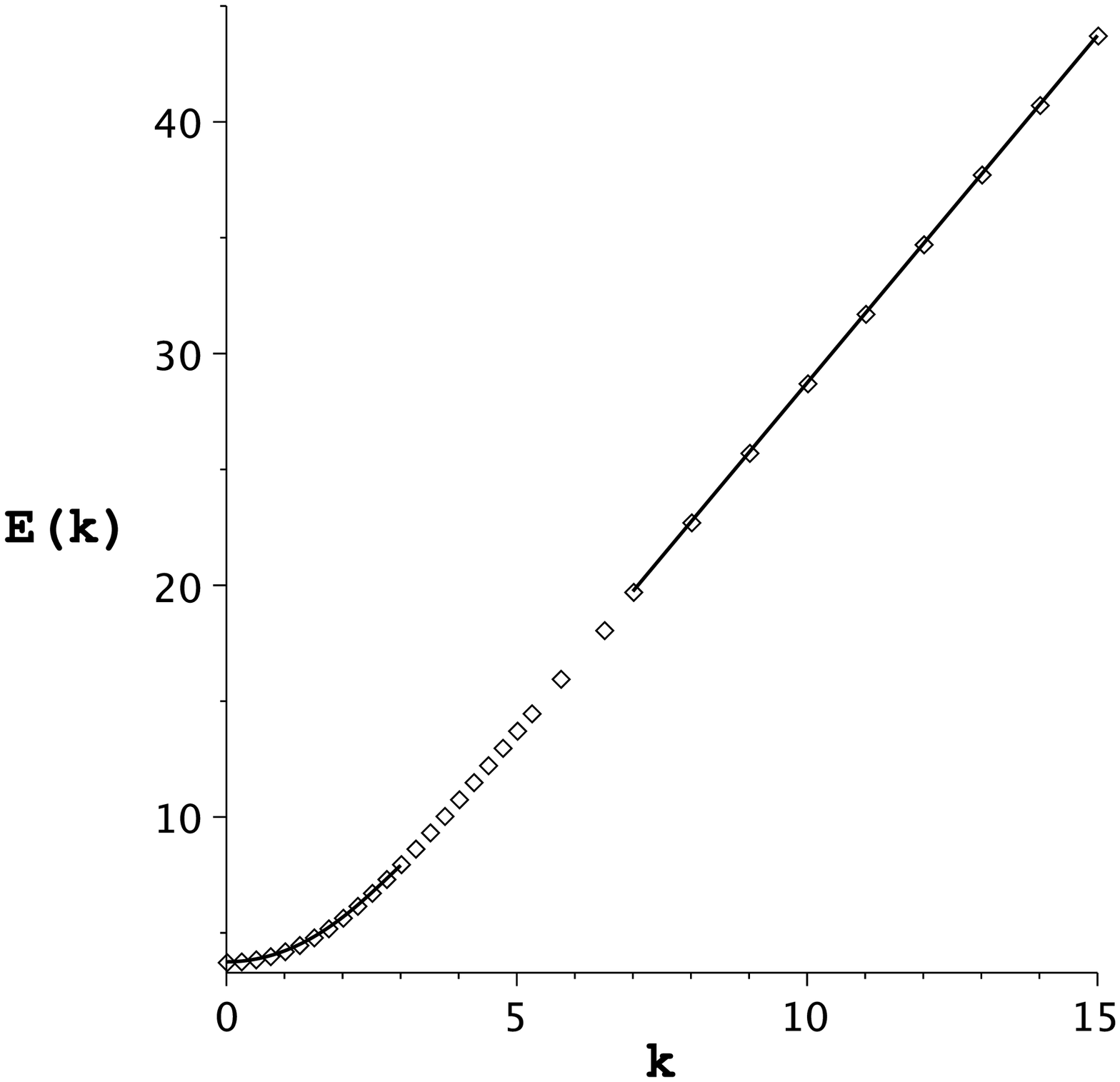}%{smallk_3.eps}%]{full_2.eps}
\includegraphics[height=7cm]{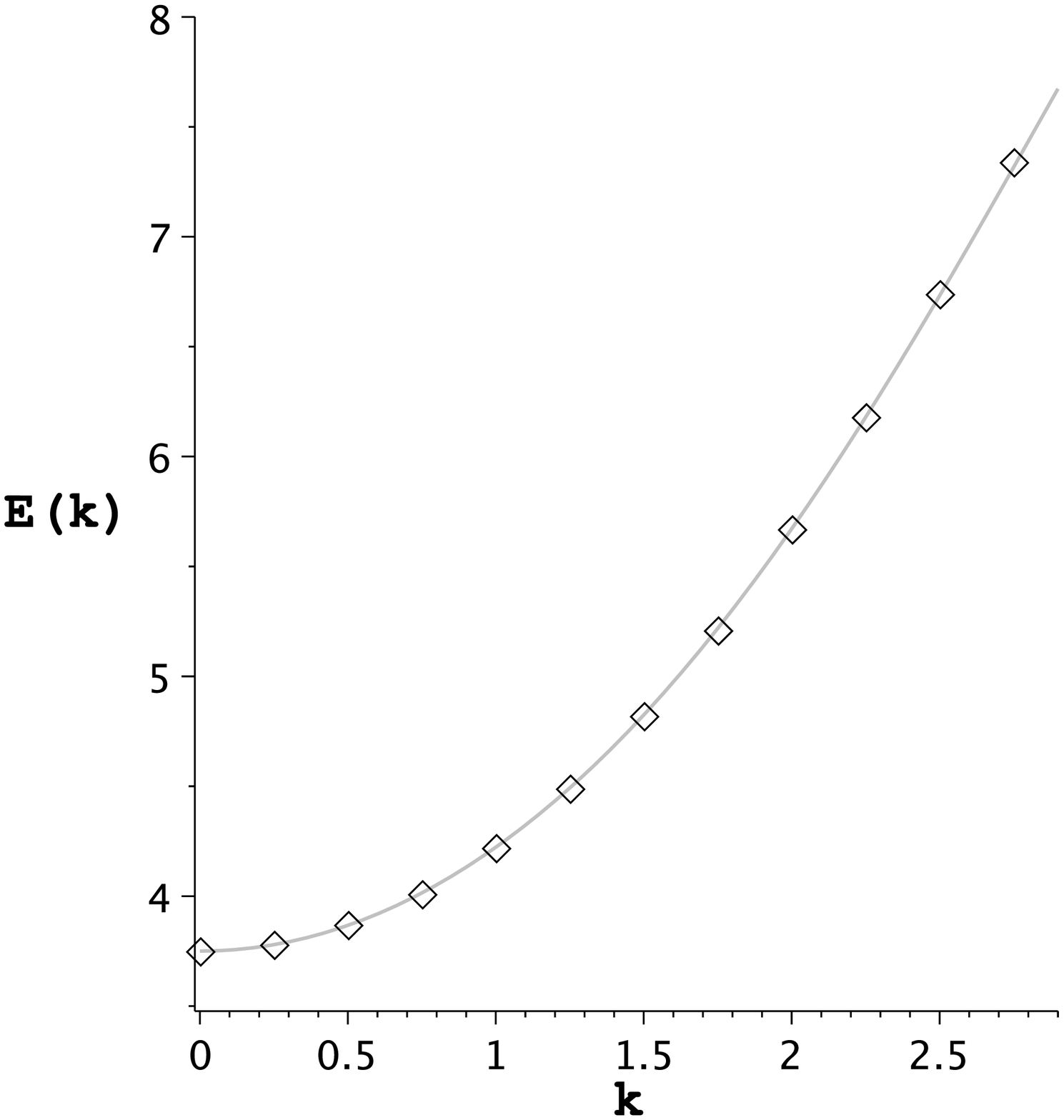}}%{fig-2.eps}%{smallk_3.eps}}
\caption{The left panel shows   $E_{\ell=1}^{ang}(s=1/2,m=0,a\w=ik)$  obtained  numerically by solving
 eq. (9) in \cite{dgrv} for $s=1/2$ and
different values of $k \in [0,15]$, together with  a least square linear fit using the large  $k$ data points,
 which gives $E=2.99843 k +$constant, in excellent agreement with the expected asymptotic expansion
 Eq. (\ref{arrivaberti}). The right panel shows the  values obtained in
this way for $0<k<2.5$. The solid line is the low frequency  approximation of  $E_{\ell=1}^{ang}(s=1/2,m=0,a\w=ik)$
in \cite{seidel,bertilong}
 to order $k^6$.}
\end{figure}

\subsection{Radial equation: reduction  to a Schr\"odinger form}

Equation (\ref{tr}) is of the form  $\Delta \ddot R + Q \dot R + (Z-E) R = 0$, dots denoting
derivatives with respect to $r$. If we introduce an integrating factor $L$,
 $\psi:=R/L$, and change the radial variable to $r^*$, where $\frac{d r^*}{dr} := \frac{1}{f}$, with
$f$  an unspecified positive definite function of $r$, (\ref{tr}) gives the following equation for $\psi$:
\begin{equation} \label{r1}
- \psi'' + \left(\frac{ f'}{f}-\frac{ 2 L'}{L} - \frac{f Q }{\Delta}  \right) \psi' +
\left(\frac{ L' f'}{L f}-\frac{ L'' }{L} - \frac{Q f L'}{L \Delta} - \frac{f^2 Z}{\Delta}  \right) \psi = \frac{f^2}{\Delta}
E \psi,
\end{equation}
where primes denote derivatives with respect to $r^*$.
By choosing $f= \sqrt{\Delta}$ (note that $\Delta$ is strictly positive for KIII and KNS)
and $L$ such that the coefficient of $\psi'$ vanishes, i.e., $L = \Delta ^ {- \frac{(2s+1)}{4}}$,
(\ref{r1}) reduces to a stationary Schr\"odinger equation with energy eigenvalue $-E$,
\begin{equation} \label{scho}
{\cal H} \psi := - \psi'' + V \psi = -E \psi.
\end{equation}
In the case $m=0, a\w=ik$, the  potential is
\begin{multline}  \label{v}
V =  - \left( \frac{\Delta \ddot L}{L} + \frac{Q \dot L}{L} + Z  \right)
= \left[\frac {r \left( {r}^{3}+r{a}^{2}+2\,M{a}^{2} \right) }{{a}^{2
} \left( {r}^{2}-2\,Mr+{a}^{2} \right) } \right]  \; k^2  +
 2s \left[\frac { \left( {r}^{3}-3\,{r}^{2}M+r{a}^{2}+ M{a}^{2} \right) }{a \left( {r}^{2}-2\,Mr+{a}^
{2} \right) }\right]  k  \\
  +  \frac{1}{4}\left[1 + \frac{(M^2- a^2)(4s^2-1)}{{r}^{2}-2\,Mr+{a}^{2}}
 \right] =: k^2 V_2 + k V_1 + V_0
\end{multline}
 We will show that the spectrum of (\ref{scho}) is entirely discrete, and use
the notation $-E_n^{rad}(s,m=0,a\w=ik), n=0,1,2,...$ for its eigenvalues, to be consistent with our previous conventions.\\
For $k$ large enough, $V$ is negative
in a region  $r_n<r<0$, and the resulting bound states lead to  unstable modes of the  Teukolsky PDE.
This is explained in detail
in the following Section.

\section{\label{proof} Unstable linear fields on Kerr' s spacetime}

The results from the previous Section can be summarized as follows: there are
solutions of the Teukolsky equations
behaving  as $e^{kt/a}$  if and only if solutions can be found of equations (\ref{ta}) and (\ref{tr}) with
the same $E$ value for $\w=ik/a$, i.e. $E^{ang}_{\ell}(s,m,\w=ik/a) = E^{rad}_{n}(s,m,\w=ik/a)$
for some $\ell$ and $n$. Since we restrict our attention to the axially symmetric case, we will drop the
$m$ index from now on, and
use (\ref{scho})-(\ref{v}) instead of (\ref{tr}).
Given that  the instability is a consequence of
the intersection of spectral lines of the angular and radial  operators for purely imaginary frequences, we need
 to gather  information on the spectra of theses operators.
Since we are interested in spotting intersections for $\w=ik/a, k \in (0,\infty)$,
we will gather information on  the asymptotic expressions
for these spectra in the limits  where $k \to 0^+$ and  $k \to \infty$.
The strategy of the proof consists in showing that in one of these limits $E^{ang}_{\ell}>E^{rad}_{0}$
whereas in the other  $E^{ang}_{\ell}<E^{rad}_{0}$, thus the intersection follows from continuity
on the spectral lines on $k$.
For the angular equation, these limits are given
  in (\ref{abajoberti}) and (\ref{arrivaberti}).
 For the radial equation, we need to work them out
and, since the analysis depends on the domain of $r$ and boundary conditions,
we will
consider separately KIII and the KNS.\\

\subsection{Unstable modes on a Kerr naked singularity}

In this Section we consider the extreme case $a^2>M^2$, for which $-\infty<r<\infty, t, \theta$ and $\phi$
are global coordinates, and $\Delta>0$ everywhere.
The choice $\frac{d r^*}{dr} := \frac{1}{f}
=1/ \sqrt{\Delta}$ made above gives an adimensional $r^*$,
\begin{equation} \label{rs}
r^* = \ln \left( \frac{r-M+\sqrt{r^2-2Mr+a^2}}{M} \right)  \simeq \begin{cases} \ln \left(\frac{2 r}{M} \right) & r \to \infty \\
\ln \left( \frac{a^2-M^2}{2 M |r| } \right) & r \to - \infty \end{cases}
\end{equation}
that grows monotonically with $r$, and  can  easily be inverted in terms of elementary functions,
\begin{equation} \label{irs}
r = \frac{M \exp(r^*)}{2} + M + \frac{M^2-a^2}{2M \exp(r^*)}.
\end{equation}
Note that (\ref{scho}) defines  a quantum mechanical problem in the entire $r^*$ line, with ${\cal H}$
in (\ref{r1}) a self adjoint operator in $L^2({\mathbb R}, dr^*)$.
The asymptotic form of the potential (\ref{v}) for large $|r^*|$ does not depend on the value of $s$, and
is given by
\begin{equation} \label{a1}
 V \sim \begin{cases}  \left( \frac{Mk}{2a} \right)^2  e^{2r^*},  & r^* \to \infty,\\
 \left( \frac{(M^2-a^2)k}{2Ma} \right)^2  e^{-2r^*}, & r^* \to -\infty. \end{cases}
 \end{equation}
From the above equation and the fact that $V$ is smooth we conclude that $V$ reaches a minimum and
${\cal H}$ in (\ref{r1}) has an entirely discrete, bounded from below spectrum $-E_n^{rad}(s,k), n=0,1,2...$
As explained above, we will need information on this spectrum
both in the $k \to 0^+$ and $k \to \infty$ limits.

\subsubsection{Asymptotic behaviour of the radial equation  spectrum as $k \to \infty$}

 The large real $k$  limit is simple to deal with, because the
behaviour of $V$ in this limit does not depend on the value of $s$, and, therefore, the analysis in
\cite{dgrv} for $s=-2$ applies with only minor modifications. The cubic polynomial
$r^3+a^2 r + 2 M a^2$ in the numerator of $V_2$ (see (\ref{v})) has a unique real root at
$r= r_n(M) <0$, thus $V_2$ is negative in the interval $r_n(M)<r<0$, and nonnegative elsewhere. Note that since
$a^2=-r_n{}^3/(2M+r_n)$, then $r_n(M)$ goes from $-M$ to $-2M$ as $a^2$ goes from $M^2$ to infinity.\\

Let $\psi$ be a properly normalized ($\langle \psi|\psi \rangle = \int_{-\infty}^{\infty} |\psi|^2 \, dr^* =1$) function
supported in the interval $(r_n(M),0)$, then
\begin{equation} \label{test1}
\langle \psi | {\cal H} | \psi \rangle =   \langle \psi | - ( \partial / \partial r^* ) ^2 |
\psi \rangle +
\sum_{j=0}^{2} k^j \langle \psi |V_j |\psi \rangle
\end{equation}
with
\begin{equation} \label{test2}
\langle \psi |V_2 |\psi \rangle =
 \int_{r_n(M)}^{0} | \psi |^2 \; V_2 \; \frac{dr}{\sqrt{\Delta}} < 0
\end{equation}
Take $k_c$ to be the largest among zero and the real roots (if any)
of
$$p(k) :=\langle \psi | - ( \partial / \partial r^* ) ^2 |
\psi \rangle + \langle \psi |V_0 |
\psi \rangle + k \langle \psi | V_1|
\psi \rangle + \frac{k^2}{2} \langle \psi | V_2|
\psi \rangle$$
(note the one half factor in the $k^2$ term!), then $p(k)<0$ for $k>k_c$ and,  if $-E_0^{rad}(s,a\w=ik)$
is the lowest  eigenvalue of ${\cal H}$ (see equation (\ref{scho})),
\begin{equation} \label{br}
 -E_{0}^{rad}(s,a\w=ik) \leq \;  \langle \psi | {\cal H} | \psi \rangle  =
\frac{k^2}{2}  \langle \psi |V_2 |\psi \rangle + p(k) \, <
\frac{k^2}{2}  \langle \psi |V_2 |\psi \rangle
 \; \; \text{ if } \; k > k_c.
\end{equation}
meaning that the absolute value of the fundamental radial level grows at least quadratically
in $k$
\begin{equation} \label{abre}
E_0^{rad}(s,a \w=ik) > \frac{k^2}{2} | \langle \psi |V_2 |\psi \rangle | \; , \; \text{ if } \; k>k_c
\end{equation}

\subsubsection{Asymptotic behaviour of the radial equation  spectrum as $k \to 0^+$}

A quick inspection to the potential (\ref{v}) gives the the minima for $k=0$ and different
spin weights:
\begin{equation}
\min \{ V(r,k=0,s), r \in {\mathbb R} \} = \begin{cases} \; \tfrac{1}{4} & , s=0 \\
\; \tfrac{1}{4} & , |s|=1/2 \\ -\tfrac{1}{2} & , |s|=1  \\ -\tfrac{7}{2}  & , |s|=2
\end{cases}
\end{equation}
A few  lengthy calculations  show however, that
\begin{equation} \label{lrb}
\lim_{k \to 0^+}  \min \{ V(r,k,s,a>M), r \in {\mathbb R} \}   = \frac{1}{4}-s^2,
\end{equation}
so that there is a discontinuity at $k=0$ for higher spin values.
Since we will be using continuity
arguments in our proof of instability, we will consider the fundamental energy $-E_0^{rad}(s,a\w=ik)$
 of the quantum Hamiltonian
in (\ref{scho})-(\ref{v}) as a function of  $k \in (0, \infty)$, for which
\begin{equation} \label{rhlb}
\lim_{k \to 0^+} -E_o^{rad}(s,a\w=ik)  > \lim_{k \to 0^+}  \min \{ V(r,k,s), r \in {\mathbb R} \}
\end{equation}
given in (\ref{lrb}).
Appendix \ref{aa}  contain the details of the calculations leading to (\ref{lrb}).

\subsubsection{ Proof of the existence of unstable modes for every $s$}

Let us gather the relevant results of the previous sections for the axially symmetric ($m=0$) modes.
For $k \to 0^+$ and $\ell=0,1,2...$
\begin{equation} \label{esta}
E_{\ell}^{ang}(s,a\w=ik)\mid_{k=0^+} = (\ell + |s|)  (\ell+|s|+1) >  s^2 - \frac{1}{4} >
E_0^{rad}(s,a\w=ik)\mid_{k=0^+},
\end{equation}
whereas for large enough positive real $k$
\begin{equation}
E_{\ell}^{ang}(s,a\w=ik) = 2 (\ell + 1) k + {\cal O} (k^0) <
\frac{k^2}{2} | \langle \psi |V_2 |\psi \rangle | < E_0^{rad}(s,a\w=ik).
\end{equation}
By continuity, we must have, {\em for every} $\ell$ and $s$, a $k_{(\ell,s)}$ such that
$E_{\ell}^{ang}(s,a\w=ik_{(\ell,s)})=E_0^{rad}(s,a\w=ik_{(\ell,s)})$. This proves that there is
 an axially symmetric unstable solution
of the Teukolsky equation for the fundamental radial level and every harmonic number $\ell$.
For higher excited radial level the arguments in \cite{dgrv} for $s=-2$ suggesting that there
also are intersections  generalize to arbitrary $s$.
In any case, we have shown that there are infinitely many unstable modes for every spin weight.
These solutions of the Teukolsky equations  decay exponentially with $|r|$
 as $|r| \to \infty$, so that they are initially  bounded, and grow exponentially in time.\\

The calculations  above can be  adapted to  perturbations in the {\em interior}
region $r< \ri:= M - \sqrt{M^2-a^2}$
of an $a \leq M$ Kerr {\em black hole}. There are some subtle
differences between the  extreme $a=M$  and
 sub-extreme $a<M$ cases, as shown  in the following sections.\\

\subsection{\label{proof2} Unstable modes on region III of an extreme Kerr black hole}

For the extreme black hole
 the solution
 of $dr^*/dr = 1/\sqrt{\Delta}$ in the interior region  $r < \ri = \ro = M$ is
 \begin{equation} \label{rse}
r^* =
-\ln \left( \frac{M-r}{M}  \right), r < \ri,
\end{equation}
with inverse
\begin{equation} \label{irse}
r = M ( 1- e^{-r^*}) , \;\;  -\infty < r^* < \infty
\end{equation}
Using  the
  integration factor $\Delta^{-\frac{2s+1}{4}}$ as before, we are led back to (\ref{scho}) and (\ref{v}), with
 $r$ given in (\ref{irse}).
  Note that
 \begin{equation} \label{av2}
V \sim \begin{cases}  4k^2 \exp(2r^*)  & , r^* \to \infty \\
                k^2 \exp(-2r^*)     & , r^* \to -\infty , \end{cases}
\end{equation}
then for any $k>0$ the   spectrum of the self-adjoint
  operator ${\cal H}$  is again fully discrete and has a lower bound.
The argument leading to (\ref{abre}) goes through in the super extreme case  without modifications,
because  the test function in (\ref{test1}) is supported in the $r<0$ region. Thus,
the fundamental energy of the radial Hamiltonian is negative and there is an
 $\ell_o$ such that $E_o^{rad}(s,a\w=ik)|_{k=0^+}<(\ell_o+|s|)(\ell_o+|s|+1)$, from where it follows
 that there is an unstable mode for every $\ell \geq \ell_o$.
The radial decay of these modes as
$r \to \ri^-$ and $r \to -\infty$ is
\begin{equation} \label{aee}
\psi \sim \begin{cases} \left(\frac{M}{M-r} \right)^ {2k-s-\frac{1}{2}} \; \exp \left[-2k \left(\frac{M}{M-r} \right) \right]
 \left( 1 + {\cal O} (\frac{M-r}{r}) \right)  & , r \to M^- \\
                   \left( \frac{M}{r}\right)^{\frac{1}{2}-2k-s} \;
 \exp \left[ \frac{rk}{M} \right]  \left( 1 + {\cal O} (M/r) \right) & , r \to -\infty  \end{cases}
\end{equation}

%Given that (\ref{av2}) is independent of $s$, the radial decay of these modes as
%$r \to \ri^-$ and $r \to -\infty$ is that given in equation (26) of \cite{dgrv}
%for $s=-2$.

\subsection{\label{proof3} Unstable modes on region III of a sub-extreme Kerr black hole}

In the sub-extreme case
  $dr^*/dr = 1/\sqrt{\Delta}$  and  $r< \ri$ give
\begin{equation} \label{urs1}
r^* = \ln \left( \frac{\ri + \ro-2r-2 \sqrt{(\ro-r)(\ri-r)}}{\ri+\ro} \right)
 \end{equation}
so that $r^*$ has an upper bound:
\begin{equation}
-\infty < r^* < \ri^* :=  \ln \left( \frac{\ro - \ri}{\ro+\ri} \right).
\end{equation}
Near the domain boundaries,
\begin{equation} \label{urs2}
r^* \simeq \begin{cases} r_i^* - \frac{2 \sqrt{\ri-r}}{\sqrt{\ro-\ri}} & , r \to \ri^- \\
\ln \left(- \left(\frac{\ro-\ri}{\ro+\ri}\right)^2 \frac{1}{4r} \right) & , r \to -\infty
\end{cases}
 \end{equation}
and
\begin{equation}  \label{sqp}
V \simeq \begin{cases}  (\nu(k)^2-\frac{1}{4})/(\ri^*-r^*)^2 & , r^* \to \ri^*{}^- \\
[k (M^2-a^2)/(2aM)]^2~\exp(-2 r^*) & , r^* \to -\infty
\end{cases}
\end{equation}
where we have defined
\begin{equation} \label{nu}
\nu(k):= 2 \sqrt{\frac{\ri}{\ro}} \left( \frac{\ro+\ri}{\ro-\ri} \right) k - s
\end{equation}

The sub-extreme case is essentially different from the extreme and super-extreme cases
because (\ref{scho}) is a
Schr\"odinger equation   {\em on the  half axis $r^*<\ri^*$}, with a
  potential that is singular at the $\ri^*$ boundary. This situation and type of
singularity is well known \cite{meetz,reed}.
Any local solution of (\ref{scho}), in particular those   which are square integrable for
  $r^*$ near   $ - \infty$, behave as
\begin{equation}
\psi \sim   a \left[(\ri^*-r^*)^{\frac{1}{2}+ \nu} + ... \right] + b  \left[ (\ri^*-r^*)^{\frac{1}{2}- \nu} + ... \right] ,  \label{zero}
\end{equation}
near the horizon.
  Thus, if  $\nu>1$,
 these are   not square integrable near the horizon, unless $b=0$,
and this is precisely the condition that
selects a discrete set of possible $E$ values as the spectrum of ${\cal H}$, and that defines de space of functions
where ${\cal H}$ is self-adjoint. This case is called {\em limit point} in \cite{reed}. It is quite different from
the {\em limit circle} case $\nu < 1$, for which  {\em for any $E$} the eigenfunction behaving properly at minus
infinity will be  square integrable
in  $r^* \in (-\infty,\ri^*)$, and a choice of boundary condition needs to be imposed to define
a set of allowed perturbations  $D_{phys}$,
in order that  ${\cal H}$ be a self-adjoint operator on $D_{phys}$ (i.e.,
$D_{phys}=D_{phys}^*$, see \cite{reed} for more details),  and thus have
 a complete set of eigenfunctions. This is done by requiring a behaviour
like (\ref{zero}) with a fixed (possibly infinite) $b/a$ ratio \cite{reed}.
Since we are ultimately interested in the large $k$ case, in view of (\ref{nu}), we do not have to deal
with this ambiguity. In any case, regardless of our choice of boundary conditions,
  the test function used in
(\ref{test1})-(\ref{test2}), being supported in a $r<0$ region,
 will belong to the chosen space of perturbations, and the argument of instability
used for the nakedly  singular Kerr spacetime will go through in the sub-extreme black
hole case if, as done for the extreme case, we restrict the harmonics to $\ell > \ell_o$
with $\ell_o$ the smallest non-negative integer satisfying $E_o^{rad}(s,a\w=ik)|_{k=0^+}<(\ell_o+|s|)(\ell_o+|s|+1)$.

\subsection{Consistency with  modal stability  outside the black hole horizon}
We would like to note that our results do not contradict the well established
modal stability of the {\em outer} stationary, region I of Kerr black holes (see
\cite{stable} en
references therein). Our arguments break down in this case since the interval where  $V_2$ is negative
lies outside
the domain of interest. \\
Consider the extreme case $a=M$, and switch to the adimensional variable
$x=r/M$, then
\begin{equation} \label{ve1}
V_{ext} = \left[ \frac{x(x+1)(x^2-x+2)}{(x-1)^2} \right] k^2 +2s \left[ \frac{x^2-2x-1}{x-1} \right] k  +\frac{1}{4}.
\end{equation}
For $s=0$,  $V_{ext}>1/4$ outside the horizon ($x>1$) and thus there is no instability.
For $s \neq 0$,  $\p V_{ext}/\p x=0$ at $x_o$ gives
\begin{equation} \label{cpve}
-\frac{k}{s} = \frac{(x_o-1)(x_o{}^2-2x_o+3)}{(x_o{}^2+1)(x_o{}^2-2x_o-1)}.
\end{equation}
The function on the r.h.s. above decreases monotonically from zero to minus infinity for $x_o \in  (1,1+\sqrt{2})$,
then from infinity to zero if $x \in (1+\sqrt{2},\infty)$. Thus, for any  $s \neq 0$, $V_{ext}$ has
a unique critical point,  which (by inserting (\ref{cpve}) in $\p^2 V_{ext}/\p x^2$) we find that is
a local, then absolute (using that $V_{ext} \to \infty$ for $x \to 1^+$ and $x \to \infty$)
minimum $V_{ext}^o$ of $V_{ext}$ in the domain of interest. This absolute minimum
 $V_{ext}^o$ can easily be seen to decrease with $k$, with a lower bound  $\frac{1}{4}-s^2$ as $k \to 0^+$.
Summarizing
\begin{equation} \label{bext}
V_{ext}(k,x,s)  > \frac{1}{4}-s^2, |s|=0,1/2,1,2; \;\; k>0, \;\; x>1.
\end{equation}
Then for the radial equation we get
$$E_0^{rad}(s,a\w=ik) < s^2-\frac{1}{4} < s^2 + |s| \leq (\ell+|s|)(\ell+|s|+1) \leq E_{\ell}^{ang}(s,a\w=ik), \;\; k>0, \ell \geq 0$$
and thus the intersection argument implying the existence of   modes that grow exponentially in time breaks down. \\
The reasoning for the sub-extreme case is similar, although the calculations are more complicated.
Instead of (\ref{cpve}) we get a two branched solution for $k$, only one of which corresponds to local minima.
The bounds (\ref{bext}) are obtained again, and thus there is no instability.

\section{\label{stm} Instabilities and time machine}

The axial Killing vector field $\zeta^a$ of Kerr spacetime ($\zeta=\p/\p \phi$ in Boyer-Lindsquit coordinates)
becomes timelike in what is called the {\em time machine} region \tm of Kerr spacetime \cite{onil},
the region where
\begin{equation} \label{tm}
%\left( \frac{r^2+a^2 \cos ^2 \theta}{\sin ^2 \theta} \right)  \zeta^c \zeta_c =
 (r^2+a^2 \cos^2 \theta) (r^2+a^2) + 2 M r a^2 \sin^2 \theta < 0.
\end{equation}
The character of the  restriction of the Teukolsky PDE  (\ref{teukopde})
to the space of  functions satisfying $\pounds_{\zeta}  \Psi = i m \Psi$) changes
from hyperbolic to elliptic within \tm.
To see this, note that the second order terms of  Teukolsky equation  are independent of $s$,
and so, for any $s$ value, they equal those of the scalar wave equation
\begin{equation}
0 = g^{a b} \nabla_a \nabla_b \Psi = \frac{1}{\sqrt{|g|}} \p_a \left( \sqrt{|g|} g^{a b} \p_b \Psi \right)
\sim g^{a b} \p_a \p_b \Psi \sim \left( g^{ab} - \frac{\zeta^a \zeta^b}{\zeta^c \zeta_c} \right) \p_a \p_b \Psi
\end{equation}
where $\sim$ means ``equal up to lower order terms'',  and $\pounds_{\zeta}  \Psi = i m \Psi$ was
used in the last step. The proof of existence of unstable modes in Teukolsky equation in the previous Section
is based on the fact that the piece $V_2$ of the potential, which is dominant for large $k$,  becomes
negative in the region $r \left( {r}^{3}+r{a}^{2}+2\,M{a}^{2} \right)$. This region is precisely
the  intersection of \tm with the equatorial plane $\theta=\pi/2$ (see (\ref{tm})). \\
The existence of instabilities seems to be related to this change of character of the
Teukolsky PDE for axial modes from hyperbolic to elliptic. To illustrate this point, consider
the simple toy model ($a,\w_o$ positive):
\begin{equation} \label{tm1}
\frac{\w_o^2 (x^2-a^2)}{4} \frac{\p^2 \Phi}{\p t ^2} - \frac{\p^2 \Phi}{\p x^2} = 0,
\end{equation}
which is hyperbolic for $|x|>a$, elliptic otherwise. Unstable solutions $\Phi(t,x)=e^{kt/a} \psi(x)$
will exist if
\begin{equation}
- \frac{\p^2 \psi}{\p x^2} + \left( \frac{k \w_o}{2} \right)^2 (x^2-a^2) \psi = 0.
\end{equation}
The above equation is that of a quantum harmonic oscillator, it has square integrable
solutions if
\begin{equation}
E = k \w_0 \left( n+ \frac{1}{2} \right)  - \left( \frac{k \w_o a}{2} \right)^2 =0, \; n=0,1,2,....
\end{equation}
Thus, there are no instabilities if $a=0$ (i.e., (\ref{tm1}) is hyperbolic everywhere). Otherwise
there will be infinitely many unstable modes, with
$$k = \tfrac{4}{\w_o a^2} \left(n + \tfrac{1}{2} \right). $$
One of the consequences of the existence of the time machine region is that it allows  to
construct a future directed timelike curve connecting {\em any ordered pair} of
 events in either KIII or KNS (see \cite{onil} and references therein).
This causes a number of difficulties when trying to define notions such as ``evolution'' and
``instability'', as discussed in the following Section.

\section{\label{rec} Unstable modes as Debye potentials}

Solutions of the transposed Teukolsky equations
of different spin weights can be used  as ``Debye potentials'' to generate
Maxwell, spinor, and linearized gravity fields \cite{reco1}. An explanation
of why this is so  was first given by Wald in \cite{reco2}, and is reviewed
in detail in appendix C of reference \cite{jhep}.
It is based on a notion of transpose of a linear differential operator ${\cal O}$
acting  on tensor fields of rank $k$, under the inner product $(U,V) := \int_M U_{a_1...a_k} V^{a_1...a_k}$,
where $M$ is the spacetime manifold and indices are raised and lowered using the metric.
The transpose is defined as usual by  $(U, {\cal O} V) = ({\cal O}^T U, V)$, and assumes
 a proper decay of the fields in the domain of ${\cal O}$, so that  integration by parts is allowed.
Suppose $f$ is the tensor field we are interested in  (e.g., the Maxwell potential  $A_b$, or the metric perturbation $h_{ab}$),
 ${\cal E} (f)=0$ the linear differential equation that it satisfies.  In  the Teukolsky formalism,
with the exception of the scalar field equation,
one does not work
with the field $f$ of interest, but with a derived field $\psi_s:= {\cal T}_s (f)$. Here ${\cal T}_s$
is a linear differential operator that projects out a null tetrad component
with spin weight $s$ of a tensor derived from $f$, e.g, a perturbed Weyl component associated
with a metric perturbation $h_{ab}$.
Teukolsky perturbative treatment can be summarized as follows \cite{reco2}: there exist  linear
differential operators ${\cal S}_s$ and ${\cal O}_s$ such that
\begin{equation} \label{rel1}
{\cal S}_s {\cal E} (f) = {\cal O}_s {\cal T}_s (f) = {\cal O}_s(\psi_s).
\end{equation}
The field equation ${\cal E} (f)=0$ then implies Teukolsky equation ${\cal O}_s(\psi_s)=0$.
The  operators $ {\cal O}_{s}$ are
defined by the left hand sides of the following equations in \cite{teuko} (refer also to
equations  (\ref{weyls})-(\ref{spin}) above):
(2.12) for $s=2$ ($\psi_2= \d \psi_0$), (2.14) for $s=-2$ ($\psi_{-2}=\delta \psi_4$),
 (3.5) for $s=1$ ($\psi_1= \phi_0$), (3.7) for $s=-1$ ($\psi_{-1}=\phi_2$), (B.4) for
$s=1/2$ ($\psi_{1/2}=\chi_0$) and (B.5) for $s=-1/2$ ($\psi_{-1/2}=\chi_1$). Using the
information  in Table I of \cite{teuko} (and calculating for the spinor case),
we find that the relation between the ${\cal O}_s$ and the operator $T_s$ in the master
Teukolsky equation (\ref{teukopde}) are
\begin{equation}
{\cal O}_s = \begin{cases} (2 \Sigma)^{-1} \circ T_s &, s \geq 0 \\
    (2 \Sigma)^{-1} \rho^{-2s} \circ T_s \circ \rho^{2s} &, s<0 \end{cases}
\end{equation}
Thus, ${\cal O}_s(\psi_s)=0$ reduces to $T_s \Psi_s =0$, where $\Psi_s=\psi_s$ for $s \geq 0$,
and $\Psi_s = \rho^{2s} \psi_s$ for $s<0$ (cf. equation (\ref{weighted})). Teukolsky master equation $T_s \Psi_s =0$
is spelled out in Boyer-Lindquist coordinates in (\ref{teukopde}).\\

As Wald noted in  \cite{reco2},  for  spinor,
Maxwell and linear gravitational
fields, ${\cal E}^T = {\cal E}$,
the transpose of ${\cal S}_s {\cal E} = {\cal O}_s {\cal T}_s$ (equation (\ref{rel1})) then gives
${\cal T}_s{}^T {\cal O}_{s}{}^T = {\cal E} {\cal S}_s{}^T$. Thus, if $\hat \psi_{s}$ is
a solution of the transposed  Teukolsky equation,
${\cal O}_{s}{}^T \hat \psi_{s} =0$, then ${\cal E}  {\cal S}_s{}^T \hat \psi_{s} =0$.
In other words, $\hat \psi_{s}$ is a ``potential'' for a solution  $f={\cal S}_s{}^T \hat \psi_{s}$
of the field equation ${\cal E} (f)=0$. \\
 A straightforward computation shows that there is a close
relation
between transpose and spin weight flip:
\begin{equation} \label{transpose}
 \rho^{2 |s|} \circ {\cal O}_{(\pm s)} {}^T \circ \rho^{-2 |s|} = {\cal O}_{(\mp s) }
\end{equation}

Solutions for the transpose of the spin weight $s$ source free
Teukolsky equations can then be readily obtained by multiplying a solution
of spin weight $-s$ times an appropriate power of $\rho$. Unstable solutions
of the Teukolsky equations will therefore produce unstable spinor, Maxwell
or gravitational fields, since the $\exp (kt/a) $ factors in the potential go through
the differential operators ${\cal S}_s{}^T$. \\

When analyzing the linear stability of a super-extreme Reissner-Nordstr\"om spacetime (or the
interior static region of a Reissner-Nordstr\"om black hole) one is
faced with the problem that the unperturbed spacetime is non globally hyperbolic due to the
timelike singular boundary \cite{dgrn}. The evolution of fields on this spacetime
is a priori not well defined, and the curvature singularity
poses the additional problem of deciding what should be considered a ``reasonable'' behaviour
for linearized  perturbations.
The way around these problems is hinted by the observation that
there is a {\em unique} choice of boundary condition at the singularity that guarantees that
the perturbed curvature invariants will not diverge faster than the unperturbed ones
 as the singularity is approached  \cite{dgrn}.
By choosing this particular boundary condition we make sure that the perturbation treatment
is self-consistent, as perturbations can be {\em uniformly} bounded on an ``initial time'' partial Cauchy surface
 $\Sigma_o$ that meets the singularity (any hypersurface orthogonal to the timelike Killing vector field).
At the same time we solve the issue of uniqueness of evolution from data given at $\Sigma_o$. Furthermore,
 this
evolution preserves the chosen boundary condition
 \cite{dgrn}. KNS, as well as   KII, also have a timelike curvature singularity, the ring singularity, located
 at $r=0, \theta=
\pi/2$ in Boyer-Lindquist coordinates. Note, however, that $r \in (-\infty, \infty)$ for
the super-extreme case ($r \in (-\infty, \ri)$ for the black hole interior), as one can
enter the $r<0$ region avoiding the singularity. The character
of the singularity and the chosen fields $\Psi_s$ is such that the separated Teukolsky equations
(\ref{ta}) and (\ref{tr}) {\em are not singular}, i.e., $r=0$ is a regular point
of the ODE (\ref{tr}), and similarly for $\theta=\pi/2$ in (\ref{ta}). When solving
(\ref{tr}), which is a second order equation, one can impose that $R(r)$ vanish as $r \to -\infty$ and $r \to \infty$
($r \to \ri$), but that leaves out any further choice, such as a selecting a specific behaviour
as $r \to 0$. This implies that, although (super extreme or black hole interior) Reissner-Nordstr\"om and Kerr
 spacetimes share some properties, such as the lack of a Cauchy
surface and the existence of a time-like singularity,
 the  issue of field propagation on those spacetimes is technically rather different,
the Kerr ring singularity
being milder. On the other hand, the causality issues are much worse in the Kerr
case. This is because, as mentioned in the Introduction, any two events in KII or KNS
 can be connected with a future directed timelike curve (in particular,
there is a closed timelike curve through {\em any} point.) There is no partial Cauchy surfaces and
thus no clear notion of ``initial time slice'' that allows to pose the stability problem as
an initial value problem.
The $t=$ constant slices are spacelike outside a compact set, and our notion of instability
is limited to the observation that  there exist  solutions
to the linear field equations behaving as $\exp (kt/a), k>0$, decaying exponentially or faster as $|r|
\to \infty$ in KNS (vanishing at the inner horizon in KIII),
and behaving ``properly'' -as defined below- near the ring singularity. \\

To check how unstable fields behave near the ring singularity of Kerr spacetime we may use unstable
solutions of the Teukolsky
master equation (proved to exist for every $s$ in the previous Section) as Debye potentials for unstable
spinor, Maxwell or gravitational fields. Note, however, that, in contrast to our previous instability results for
the Reissner-Nordstr\"om or naked Schwarzschild case case \cite{dottigleiser,dgrn}, we lack of {\em explicit}
expressions for the unstable solutions
of the Teukolsky master equations. We can still get information on the behaviour of unstable fields
near the ring singularity by analyzing
the Frobenius series solutions of the ODE (\ref{tr}) near $r=0$ and the ODE (\ref{ta}) near
$\theta=\pi/2$. These series, however, are
 independent of the stable or unstable character of the solution, since $\w$ does not show up
at leading order. Thus, whatever criterion we adopt to disregard field solutions from
Debye potentials based
on their  behaviour near the singularity, it will overrule {\em every} field (unstable or not)
that can be constructed using the potential method outlined above. \\

\subsection{Maxwell fields}

For Maxwell fields $s=\pm1$, and the operators and fields in (\ref{rel1}) are
\begin{equation}
f=A_b , \;\; [{\cal E}(A_b)]_a = \nabla^c \nabla_c A_a - \nabla^c \nabla_a A_c,
\end{equation}
and, for $s=1$,
\begin{eqnarray} \label{sdmf}
{\cal T}_1(A_b) &=& l^a  m^b (\nabla_a A_b - \nabla_b A_a)\\
S_1(J_a) &=& \tfrac{1}{2} (\d-\b-\bar \a - 2 \tau+\bar \pi) (j_c l^c) - \tfrac{1}{2}(D-2 \rho-\bar \rho) (j_c m^c)\\
{\cal O}_1(\psi_1) &=& (D-2 \rho- \bar \rho)  (\Delta +\mu - 2 \gamma) \psi_1 - (\d -\b
-\bar \alpha -2 \tau+ \bar \pi) (\bar \d + \pi - 2 \alpha) \psi_1,
\end{eqnarray}
where the standard null tetrad formulation notation \cite{np,teuko} is used ($D,\Delta, \d, \bar \d$
are derivatives along the tetrad vectors, the other symbols represent spin coefficients.)
Since ${\cal T}_1$ projects out  a self-dual piece of $F_{ab}$ (see (\ref{sdmf})),
 the complex potential
${\cal S}_1{}^T \hat \psi_{1}$ constructed from a solution ${\cal O}_1{}^T (\hat \psi_1)=0$ of the transpose
 Teukolsky equation \cite{reco2},
\begin{equation} \label{recomax}
[{\cal S}_1{}^T \hat \psi_{1}]_b= [-l_b ( \d + 2 \b + \tau) + m_b (D + \rho) ] \hat \psi_1,
\end{equation}
will produce a self dual Maxwell
field $G_{ab}=F_{ab}+i\, {}^*F_{ab}$ \cite{reco2}. In fact, the easiest way to check that
the exterior derivative $G_{ab}$ of the potential (\ref{recomax}) satisfies the source-free Maxwell equations
is by checking that it is self-dual, which amounts  to checking that
the contractions of $G_{ab}$ with any of
the three anti-self-dual two-forms  obtained by complex conjugation of (\ref{sdb}) vanishes as
a consequence of ${\cal O}_1{}^T (\hat \psi_1)=0$.

To evaluate  the strength of the real Maxwell
field  $F_{ab}$
near the ring singularity,
we  compute the algebraic invariants $I_1 = F_{ab} F^{ab}, I_2 = F_{ab}{}^*F^{ab}$ (any other algebraic
invariant will be a polynomial on these). Note that, since
$\tfrac{1}{2} G_{ab}G^{ab} = I_1 + i I_2 =: I$, we can compute the invariants of $F_{ab}$
more efficiently without even taking the real part of $G_{ab}$.
 For generic separable solutions $\hat \Psi_1 =
e^{i \w t} R(r) S(\theta)$ of the $s=-1$ Teukolsky mater equation,  (\ref{recomax}) gives
a field  whose invariants
admit an expression that can be simplified near the ring singularity  by applying
 iteratively  the equation $
T_{-1} (\hat \psi_1) = 0$, to
\begin{equation} \label{gmd}
I \simeq {\frac{2
\left( -iS(\theta)a \frac{dR(r)}{d r}+R(r) \frac{dS(\theta)}{d\theta} \right) ^{2}
  {{\rm e}^{2i \w t/a}} }{ \left( {r}+ i {a}  \cos \left( \theta \right)  \right) ^{4}}}.
\end{equation}
 As already explained, this  leading order
term  (omiting the $\exp(2i \w t/a)$ factor)
  will be the same for any complex $\w$. This behaviour near the ring singularity is universal,
and thus independent of the un/stable character of the field.\\
To evaluate weather or not the above divergency is ``reasonable'', we may compare with the static
Maxwell field on Kerr that we get from the Kerr-Newman solution
\begin{equation} \label{statmax}
F = dA, \;\; A = \frac{Qr}{\Sigma} \left( dt - a \sin ^2 (\theta) \,  d\phi \right)
\end{equation}
Note that since the Kerr-Newman metric is quadratic in $Q$,
 this field is a {\em first order in $Q$} solution of Maxwell equations on  a fixed Kerr
metric and,
 being Maxwell equations linear, (\ref{statmax})
 is also an {\em exact} solution on the Kerr background.
For
this static field, a straightforward calculation shows that
\begin{equation}
I_{static} = \frac{-Q^2}{2 (r - i a \cos \theta ) ^4}
\end{equation}
which exhibits  the same degree of divergency as (\ref{gmd}), the latter  being possibly even milder
along selected directions, or for some particular solutions.
Note that the unstable solutions of the $s=-1$ Teukolsky master equation, which  evolve as
$e^{kt/a}$,  decay in KNS as $e^{-k|r|/a}$ for large $|r|$, as opposed to the
slow, $r^{-4}$ decay of  the invariants of the static field abov. The unstable modes of  KIII
 decay exponentially as $r \to \infty$, and as a power of $r- \ri$ towards
the inner horizon.\\
 In conclusion, we have shown that there are solutions
of the Maxwell equations that behave  in a similar way as the static field from the Kerr-Newman solution
near the ring singularity, decay much
faster away of along in spacelike directions, and grow exponentially with time. \\

\section{Conclusions}

We have proved the existence of instabilities in both Kerr naked singularities (KNS) and 
the region beyond the inner horizon  of sub extreme and extreme Kerr black holes (KIII).
 Our notion of instability is given by 
the existence of solutions  of linear massless field equations that behave as $e^{kt}, k>0$
with a fast decay to zero as $|r| \to \infty$ in the case of KNS (as $r \to -\infty$ and $r \to \ri^-$ 
in the case of KIII), where $\{ t,r,\theta, \phi\}$ are Boyer-Lindquist coordinates. 
We have shown that there exist massless scalar fields, Weyl spinors, Maxwell fields and linear gravity perturbations 
with these properties.\\
Since KIII and KNS do not admit a partial Cauchy surface, there is no natural notion 
of evolution from initial data, and therefore of instability in the usual way (bounded data grows unbounded).
However, these spacetimes  are time orientable, since the never vanishing  vector field 
$$V = (r^2+ a^2) \frac{\p}{\p t} + a \frac{\p}{\p \phi}  $$
is everywhere timelike. The integral lines of $V$ give a congruence of timelike curves filling the entire 
KIII (KNS) space which may  be regarded
as worldlines of (accelerated) observers. Since the unstable fields we have found are axially symmetric
(independent of $\phi$), they grow boundless along these curves. That a congruence of observers 
measure a boundless growth of these linear fields (more properly, of any scalar made out from them) is 
to us  an appropriate notion of instability in spacetimes lacking a partial Cauchy surface (for
examples of evolution and stability notion in non globally hyperbolic spacetimes admitting a partial Cauchy surface see 
the Schwarzschild and Reissner-Nordstr\"om cases \cite{dottigleiser}, \cite{dgrn}).)\\
A modification of the KNS metric based on  alternative string motivated 
theories, or just the excision of a region around the ring singularity, gives rise 
to ``superspinars'', compact rotating objects violating the black hole $a \leq M$ bound \cite{gh}.
The gravitational stability of these objects (more concretely, the region $r>r_0$ of KNS, assuming different 
boundary conditions at $r=r_0$) was studied in detail in  \cite{pani} and references therein. In Section 3.B of that paper,  
the superspinar gravitational instabilities results are confronted with those in our works 
\cite{doglepu} and \cite{dgrv}, generalized in the present paper.
 This is done by studying  the  $r_0 \to -\infty$ limit 
of unstable axially symmetric perturbations of a superspinar with a perfectly reflecting ``string horizon'' at $r_0 \to  -\infty$.
This gives a result that agrees perfectly with our previous numerical calculations in \cite{dottigleiser}.\\
The mathematical origin of the instability in both cases (KNS and superspinars) is, however, quite different. In the KNS 
case it is due to the negative portion of the effective potential (\ref{v}) for small $r<0$ values, 
a region that is usually excised in a superspinar. The superspinar stability problem 
reduces also to a Schr\"odinger equation problem. However, in the KNS case, the 
 Schr\"odinger equation  is posed on the entire real axis, whereas in the 
superspinar case is a quantum mechanics problem on a half axis (which may well exclude
the region where our effective potential is negative), with non trivial boundary conditions 
that are responsible of the instability. For KIII in the sub-extreme case we have also reduced the field
equations to a  QM problem 
on a half-axis, however, the boundary conditions are trivial, and the negative portion 
of the effective potential belongs to this half axis.\\
 We should also comment on  the 
connection found  in \cite{pani} (see also \cite{otras}) 
between circular geodesics with negative energies and superspinar instabilities. 
We note that those results do not seem to apply in a straightforward way to our case, since the arguments in \cite{pani} apply 
for $\ell=m >>1$ modes whereas unstable modes in this work have $m=0$. We should also stress that, contrary to what is
 found in \cite{pani} the unstable modes
whose existence is proved in this work exist for any value of $a/M$ as long as $a/M>1$.

\section*{Acknowledgments}

This work was supported in part by grants from CONICET
and Universidad Nacional de C\'ordoba.  GD is supported by CONICET and Universidad
Nacional de C\'ordoba. RJG is partially supported by
CONICET,  IFRS is a fellow of CONICET.

\appendix

\section{Lower bounds to the radial potentials for the KNS as $k \to 0^+$} \label{aa}

This Appendix gathers the calculations leading to the bounds given in (\ref{lrb}) for
the global minima of the potential (\ref{v}) in the $k \to 0^+$ limit.
These are used in Section \ref{proof} to obtain a bound if the radial spectrum in this limit.

\subsubsection{Case $s=0$}
 Introducing the adimensional variables $x=r/M, \a=a/M$, the critical points of the potential (\ref{v}) for $s=0$ are given by
\begin{equation} \label{cps0}
4 k^2 \;(\a^2+x^2)(x^3-3x^2+\a^2x+\a^2) + \a^2(\a^2-1) (1-x)=0
\end{equation}
For $k$ small enough, three out of the five roots are real, and we order them as $x_1<x_2<x_3$.
Given the asymptotic behaviour (\ref{a1}), it is clear that the absolute minimum of $V$
is reached at some of these points ($x_1$ or $x_3$ if there are no inflection points).
 Inspection of the
numerical solutions of (\ref{cps0}) for increasingly smaller $k$ values suggests that
$x_1 \to -\infty$, $x_2 \to 1$  and $x_3 \to \infty$ as $k \to 0^+$. Guided by this
observation we propose a solution of (\ref{cps0}) in the form of a power series in $k$ taking
the value $x_2=1$ at $k=0$, and obtain
by iteration the successive corrections in increasing powers of $k$, the result being,
\begin{equation} \label{x2}
x_2 = 1 + \frac{8(1+\a^2)}{\a^2} k ^2 + \frac{32 (3 \a^4- 5\a^2+ \a^6-7)}{\a^4(\a^2-1)} k ^4 +   {\cal O} (k^6)
\end{equation}
Then we assume $x=x_1$ in (\ref{cps0}), solve for $k$, and expand the resulting expression
for $x_1 \to \infty$ (which we know that corresponds to taking $k \to 0^+$) to extract the leading order behaviour of
the relation between $x_1$ and $k$, which is $x_1 = -((\a^2-1)/4)^{1/4} k^{-1/2}$.
Inserting this expression plus a  correction (to be obtained) back into
(\ref{cps0}),  we can iteratively get as  many  higher order terms as we wish. The first few of them are
\begin{equation}
x_1 = -\left(\frac{\a^2-1}{4}\right)^{1/4} k^{-1/2} + \frac{1}{2} + \left[\frac{\sqrt{2} (4 \a^2-7)}{8 (\a^2-1)^{1/4}}\right]
\;  k ^{1/2} + {\cal O} (k).
\end{equation}
Proceeding in a similar way leads to
\begin{equation}
x_3 = \left(\frac{\a^2-1}{4}\right)^{1/4} k^{-1/2} + \frac{1}{2} - \left[\frac{\sqrt{2} (4 \a^2-7)}{8 (\a^2-1)^{1/4}}\right]
\;  k ^{1/2} + {\cal O} (k)
\end{equation}
The value of the potential at these points are $V_1 = 1/4 +  {\cal O} (k), V_2 = 1/2 +  {\cal O} (k^2), V_3 =
1/4 +  {\cal O} (k)$. It then follows that
\begin{equation}
\lim_{k \to 0^+}  \min \{ V(r,k,s=0,\a), r \in {\mathbb R} \} = 1/4,
\end{equation}
in agreement with (\ref{lrb})

\subsubsection{Case $|s|= 1/2$}
For $s=1/2$, there are three real critical points $x_1<x_2<x_3$ ($x_1$ and $x_3$ are local minima, $x_2$
a maximum)  only if $k$ is bigger than an $a$ dependent
critical value, below which $x_2$ and $x_3$  coalesce
into a single (inflection) point. In any case, the absolute minimum is reached at $x_1$, for which
\begin{equation} \nonumber
 x_1 = -\frac{\a}{2k}-1 -\frac{4k}{\a}- \frac{16(\a^2-4)}{\a^2} k^2 + {\cal O}(k^3)\;\;,   V_1 =
-\frac{2}{\a} k + \frac{7}{\a^2} k^2 +  {\cal O}(k^3) \\
\end{equation}
For $s=-1/2$, the analysis is similar, with $x_1 \to x_2$ and a global minimum at $x_3 \to \infty$ as $k \to 0^+$:
\begin{equation} \nonumber
 x_3 = \frac{\a}{2k}-1 +\frac{4k}{\a}- \frac{16(\a^2-4)}{\a^2} k^2 + {\cal O}(k^3)\;\;,  V_3 =
\frac{2}{\a} k + \frac{7}{\a^2} k^2 +  {\cal O}(k^3) \\
\end{equation}
 It  follows that
\begin{equation}
\lim_{k \to 0^+}  \min \{ V(r,k,s=\pm 1/2,\a), r \in {\mathbb R} \} = 0,
\end{equation}
in agreement with (\ref{lrb})

\subsubsection{Case $|s|= 1$}
For $s=\pm 1$, the critical points and potential values at these points are

\begin{align}  \nonumber
 x_1 &= 1\mp \frac{4}{3}\,{\frac { \left( {\alpha}^{2}-3 \right) k}{\alpha}}-{\frac
{8}{9}}\,{\frac { \left( 11\,{\alpha}^{2}-21 \right) {k}^{2}}{{
\alpha}^{2}}} + {\cal O}(k ^3),
 \\ \nonumber
 x_2 &= \mp {\frac {\alpha}{k}}-1\pm \frac{1}{4}\,{\frac { \left( -11+3\,{\alpha}^{2}
 \right) k}{\alpha}}-\frac{1}{2}\,{\frac { \left( -41+17\,{\alpha}^{2}
 \right) {k}^{2}}{{\alpha}^{2}}}+ {\cal O}(k ^3)\\  \nonumber
x_3 &= \frac{1}{2}\,\sqrt [3]{6}\sqrt [3]{\mp {\frac { \left( -1+{\alpha}^{2} \right)
\alpha}{k}}}+1+ {\cal O}(k ^{1/3}) 
\end{align}
the values of the potential at these points being
\begin{align} \nonumber
V_1 &= -\frac{1}{2} \pm \frac{4}{\a} k + {\cal O}(k ^2)\\ \nonumber
V_2 &= -\frac{3}{4} \mp \frac{4}{\a} k + {\cal O}(k ^2)\\
V_3 &= \frac{1}{4} - \frac{3^{4/3}}{2^{2/3}} \frac{\a ^2 - 1}{(\a (\a ^2 - 1))^{2/3}} k ^{2/3}+ {\cal O}(k ^{4/3}).  \nonumber
\end{align}
 Thus
\begin{equation}
\lim_{k \to 0^+}  \min \{ V(r,k,s=\pm 1,\a), r \in {\mathbb R} \} = -3/4,
\end{equation}
in agreement with (\ref{lrb}).

\subsubsection{Case $|s|=2$}
For $s=\pm 2$ the critical points and potential values are
\begin{align}  \nonumber
x_1 &= 1 \mp {\frac {8}{15}}\,{\frac { \left( {\alpha}^{2}-3 \right) k}{
\alpha}}-{\frac {8}{225}}\,{\frac { \left( 47\,{\alpha}^{2}-81
 \right) {k}^{2}}{{\alpha}^{2}}} + {\cal O} (k^3) \\ \nonumber
 x_2 &= \mp 2\,{\frac {\alpha}{k}}-1 \pm \frac{1}{32}\,{\frac { \left( -47+15\,{\alpha}^
{2} \right) k}{\alpha}}-\frac{1}{32}\,{\frac { \left( -173+77\,{\alpha}^{
2} \right) {k}^{2}}{{\alpha}^{2}}}+ {\cal O}(k^3)\\ \nonumber
x_3 &=  \left( \pm \frac{15}{8}\, \left( 1-\,{
\alpha}^{2} \right) \alpha \right)^{1/3}{k}^{-1/3} +1+ {\cal O }(k ^{1/3})\;\;,
\end{align}
and
and the values of $V$ at these points are
\begin{align} \nonumber
 V_1 &= -\frac{7}{2} \pm \frac{8}{\alpha}k +  {\cal O} (k^2); \\ \nonumber
V_2 &= -\frac{15}{4}  \mp \frac{8}{\alpha} k + {\cal O} (k ^2) \\ \nonumber
 V_3 &=  \frac{1}{4} - \frac{3\, 15^{1/3}(\alpha ^2 -1)}{(\alpha(\alpha ^2 -1))^{2/3}}k ^{2/3} + {\cal O}(k ^{4/3}). %  \nonumber
\end{align}
It follows that
\begin{equation}
\lim_{k \to 0^+}  \min \{ V(r,k,s=0,\a), r \in {\mathbb R} \} = -15/4.
\end{equation}
This completes the verification of  (\ref{lrb}).

\end{document}